\documentstyle[amsmath,amssymb,graphicx]{article}

\def\be{\begin{eqnarray}}
\def\ee{\end{eqnarray}}

\def\tr{{\rm tr}\,}

\hoffset= - 1.0in         % switch off for draft style
\title{{\bf On "Dotsenko-Fateev" representation of the toric conformal blocks}
\vspace{.2cm}}
\author{{\bf A.Mironov}\footnote{ {\small {\it
Lebedev Physics Institute} and {\it ITEP, Moscow, Russia}};
mironov@itep.ru; mironov@lpi.ru}, {\bf A.Morozov}\thanks{{\small
{\it ITEP, Moscow, Russia} and
{\it Laboratoire de Mathematiques et Physique Theorique,
CNRS-UMR 6083,
Universite Francois Rabelais de Tours,
France}}; morozov@itep.ru} \ and {\bf
Sh.Shakirov}\thanks{{\small {\it ITEP, Moscow, Russia} and {\it
MIPT, Dolgoprudny, Russia}}; shakirov@itep.ru}\date{ }}

\begin{document}

\maketitle

\vspace{-6.0cm}

\begin{center}
\hfill FIAN/TD-12/10\\
\hfill ITEP/TH-37/10\\
\end{center}

\vspace{4cm}

\begin{abstract}
We demonstrate that the recent ansatz of \cite{MY}, inspired by the
original remark of R.Dijkgraaf and C.Vafa, reproduces the toric
conformal blocks in the same sense that the spherical blocks are
given by the integral representation of \cite{MMS1,MMS2} with a peculiar
choice of open integration contours for screening insertions. In
other words, we provide some evidence that the toric conformal
blocks are reproduced by appropriate $\beta$-ensembles not only in
the large-$N$ limit, but also at finite $N$. The check is explicitly
performed at the first two levels for the 1-point toric functions.
Generalizations to higher genera are briefly discussed.
\end{abstract}

\section{Introduction}

In \cite{MMS1,MMS2,MMMor} we suggested a matrix model type
($\beta$-ensemble) representation of conformal blocks, which we
called "Dotsenko-Fateev representation". It is based on the old idea
of "conformal matrix models" \cite{MMMar,KMMMP,DV} and it differs
from the original Dotsenko-Fateev formulas for conformal blocks in
minimal models \cite{DF,Bos,Fel} by substitution of the closed
integration contours of screening operators by a carefully chosen
open (non-closed) contours with a single (rather than two) screening
operator. This integral representation of conformal blocks is
believed to be helpful in study of the still mysterious AGT relation
\cite{AGTfirst}-\cite{AGTlast}, and it has been further explored in
a number of interesting papers \cite{Ito}-\cite{MY}.

In particular, in an very recent ref.\cite{MY}
a generalization from spherical to toric conformal
blocks was put forward in the large $N$ limit,
based on the original
remark in \cite{DV}. Similar generalization
at finite $N$, i.e,
in the same sense as the spherical block was reproduced
in refs.\cite{MMS1,MMS2,MMMor}, was proposed in ref.\cite{MMS2}, s.5,
for a particular class of toric conformal blocks.
The results of \cite{DV,MY} imply how this generalization
can be done for the generic toric conformal block at
finite $N$.
The goal of the present paper is to demonstrate that the suggestion of
\cite{DV,MY} indeed, works at finite $N$.
We show explicitly that the 1-point toric function (conformal block)
\be
B\big( q \big) = 1 + B_1 q + B_2 q^2 + \ldots
\label{cobloexp}
\ee
with external dimension $\Delta_{ext}$, internal dimension $\Delta$, central charge $c$ and coefficients
\be
B_1 = \dfrac{\Delta_{ext}(\Delta_{ext}-1)}{2\Delta} + 1
\label{B1}
\ee
\begin{align}
\nonumber B_2 \ = \ & \dfrac{1}{4 \Delta (c+2 c \Delta-10 \Delta+16 \Delta^2)} \times \Big(
(8 \Delta+c) \Delta_{ext}^4+(-64 \Delta-2 c) \Delta_{ext}^3+(8 c \Delta+3 c+56 \Delta+128 \Delta^2) \Delta_{ext}^2+ \emph{} \\ \nonumber & \\ & \emph{} +(-2 c-8 c \Delta-128 \Delta^2) \Delta_{ext}-80 \Delta^2+128 \Delta^3+8 c \Delta+16 c \Delta^2 \Big)
\label{B2}
\end{align}
is reproduced, at least at the first two orders of $q$-expansion,
by the generalized matrix model of the form (cf. \cite[formula (63)]{MMS2})
\be
Z_{DF}\big( q \big) = \int\limits_{0}^{2\pi} dz_1 \ldots
\int\limits_{0}^{2\pi} dz_N \prod\limits_{i < j} \theta_*(z_i - z_j)^{2\beta}
\prod\limits_{i} \theta_*(z_i)^{2\mu} \prod\limits_{i} e^{I A z_i} = {\rm const} \cdot \Big( 1 + J_1 q + J_2 q^2 + \ldots \Big)
\label{Ztor}
\ee
where $\theta_*(z)$ is (the exponent of) the holomorphic Green function of free fields on a torus:
\be
\theta_*(z) = \sum\limits_{n = 0}^{\infty} (-1)^n q^{n(n+1)/2} \sin\frac{(2n+1)z}{2} = \sin\frac{z}{2} - q \sin\frac{3z}{2} + q^3 \sin\frac{5z}{2} +\ldots
\ee
The partition function $Z_{DF}(q)$ can be viewed as a two-fold generalization of conventional eigenvalue matrix models:
first, the ordinary differences $z_i - z_j$ are substituted by $\theta_*(z_i - z_j)$, second, they are raised to arbitrary powers
$\beta$. These types of generalizations are often called elliptic- and $\beta$-deformations, respectively.
For reader's convenience, we still use the term (generalized) matrix model for $Z_{DF}(q)$, instead of the more precise term
"elliptic $\beta$-ensemble".

As explained in \cite{MMS2}, in the spherical case such matrix models naturally appear as free-field correlators with insertions
of $N$ screening operators, integrated over their positions $z_1, \ldots, z_N$. The main new ingredient for the torus
(except for an obvious modification of the Green function) is introduction of an additional parameter $A$ into the partition function.
Our aim is to check, perturbatively at levels one and two, the following statement:
\be
\boxed{
\left( \prod\limits_{n = 1}^{\infty} (1 - q^n) \right)^{\nu} \times \Big( 1 + j_1 q + j_2 q^2 + \ldots \Big) = \Big( 1 + B_1 q + B_2 q^2 + \ldots \Big)
}
\label{MainIdent}
\ee
Here $j_k = J_k\big(A,\mu = -\beta N,\beta,N\big)$ are the "on-shell" coefficients of the partition function, i.e. those taken at
particular values of $\mu$, satisfying the momentum conservation law $\mu + \beta N = 0$. We check the correspondence and
find the precise relation between the parameters $A$, $\beta$, $N$ of the Dotsenko-Fateev partition function and parameters
$\Delta_{ext}$, $\Delta$, $c$ of the conformal block (as well as the parameter $\nu$, the power of the $U(1)$-factor). Our result in
the 1-point sector is
\be
\boxed{
\begin{array}{ccc}
\Delta = \dfrac{A^2 - (\beta - 1)^2}{4\beta}, \ \ \ \ \Delta_{ext} = \beta N^2 + \beta N - N \\
\\
\nu = 3 \Delta_{ext} + 3 N - 1 \\
\\
c = 1 - 6 \left( \sqrt{\beta} - \dfrac{1}{\sqrt{\beta}} \right)^2 \\
\\
\end{array}
}
\ee
\smallskip\\
The present state of knowledge, therefore, can be summarized as follows.
It is checked that Dotsenko-Fateev type $\beta$-ensembles
with certain choices of integration contours describe the $n$-point spherical,
and $1$-point toric conformal blocks exactly (at finite $N$).
A similar check for the $n$-point toric functions remains to be done,
though there are few doubts in the validity of the conjecture at genus one.
Much more interesting and yet obscure is the generalization to higher genera.
This generalization is briefly discussed in s.\ref{conc}
at the end of the present paper.

\section{Partition function}
\subsection{Method of calculation}

To test the relation and find the correspondence between parameters,
one needs a method of calculation of expressions $J_k$. We exploit
the method of analytical continuation: to find a particular $J_k$,
one expands the integrand (product of the theta-functions
$\theta_*(z_i - z_j)$ and $\theta_*(z_i)$) in integer powers of $q$, and use
the rule
\be
\int\limits_{0}^{2\pi} e^{iK z + iA z} dz = \dfrac{i(1
- e^{2 \pi i A})}{A + K}
\ee
for integer $K$ to integrate all the
particular terms in the expansion. For natural $N$, $\beta$ and
$\mu$, each finite order of the expansion contains a finite number
of terms. Therefore, for natural values of parameters, this method
easily allows one to find any particular $J_k$. Given these values
at natural numbers, one can then restore the whole function by (the
simplest variant of) analytical continuation: just assuming the
dependence at non-natural values of parameters is given by a rational function.
This method of
calculation was suggested in \cite{MMS2},
and we use it in the present paper as well.

\subsection{Level 1}

At level 1, for natural $N$, $\beta$ and $\mu$, one finds
\begin{center}$
J_1\big(A,\mu,\beta,N\big) = \Big[ \beta^3 N^4+(-5 \beta^3+4 \mu \beta^2+2 \beta^2) N^3+(7 \beta^3+\beta+3 \beta \mu^2-3 A^2 \beta+6 \beta \mu-6 \beta^2-12 \mu \beta^2) N^2+(-6 A^2 \mu-2 \mu^3+3 A^2 \beta+8 \mu \beta^2-6 \beta \mu-\beta+4 \beta^2-3 \beta^3-3 \beta \mu^2+2 \mu) N \Big]\Big[ (A + \mu + \beta N - \beta + 1)(A - \mu - \beta N + \beta - 1) \Big]^{-1}
$\smallskip\\ \end{center}
At particular value of $\mu = -\beta N$ this turns into
\be
j_1\big(A,\beta,N\big) = \dfrac{\beta N(N+1) (2 \beta^2 N^2-4 \beta N+2 \beta^2 N+4 \beta+3 A^2-1-3 \beta^2)}{(A - \beta + 1)(A + \beta - 1)}
\label{J1}
\ee
\smallskip\\
Comparing this expression with eq. (\ref{B1}), one finds $B_1 = j_1 - \nu$
provided that
\be
\Delta_{ext} = \beta N^2 + \beta N - N, \ \ \ \Delta = \dfrac{A^2 - (\beta - 1)^2}{4\beta}, \ \ \ \nu = 3 \Delta_{ext} + 3 N - 1
\ee
\smallskip\\
This can be regarded as establishing the correspondence between parameters
in (and simultaneously a check of) identity (\ref{MainIdent}).
Let us proceed to level 2, to further check relation (\ref{MainIdent}).

\subsection{Level 2}

At level 2, the relation being tested, eq. (\ref{MainIdent}) turns into
\be
B_2 = j_2 - (3 \Delta_{ext} + 3 N - 1) (j_1 + 1) + \dfrac{(3 \Delta_{ext} + 3 N - 1)(3 \Delta_{ext} + 3 N - 2)}{2}
\label{Lvl2}
\ee
\smallskip\\
where $B_2$ and $j_1$ are given by (\ref{B2}) and (\ref{J1}), respectively.
As can be deduced from general considerations, the function $j_2$ is a polynomial in
$N$ of degree 8; thus, to determine this polynomial unambiguously, one needs to
calculate its values at 9 distinct natural values of $N$: say, from 1 to 9.
Unfortunately, a direct computation of the partition function at these values of $N$
is hardly feasible: the corresponding computer programs require too much time and
memory for a successful run. For this reason, we select a slightly different way
to test the relation at level 2: namely, we find $j_2$ from (\ref{Lvl2}) and then
test this \emph{prediction} at many particular values of $N$ and $\beta$.

Assuming the conventional dependence
\be
c = 1 - 6 \left( \sqrt{\beta} - \dfrac{1}{\sqrt{\beta}} \right)^2
\ee
\smallskip\\
the relation (\ref{Lvl2}) implies the following \emph{prediction} for $j_2$:
\begin{center}
$j^{(theor)}_2 = \beta N(N+1)(-12+45 A^2-76 \beta N+120 \beta-222 A^2 \beta+94 \beta^2 N^2-209 \beta^3 N^2+366 \beta^2 N-745 \beta^3 N+176 \beta^3 N^3-417 \beta^2+678 \beta^3+121 A^2 \beta N-32 \beta^2 N^3-4 \beta N^2-48 \beta^3 N^4+25 \beta N^2 A^2-42 A^4-561 \beta^4+411 A^2 \beta^2+148 \beta^4 N^4-480 \beta^4 N^3+108 \beta^4 N^2+864 \beta^4 N-136 \beta^5 N^4+624 \beta^5 N^3+39 \beta^5 N^2-601 \beta^5 N-16 N^6 \beta^5+64 N^5 \beta^4-184 N^5 \beta^5+46 N^4 \beta^6+34 N^6 \beta^6+166 N^5 \beta^6-16 N^6 \beta^7-48 N^5 \beta^7+228 \beta^5+331 A^2 \beta^3 N^2+559 A^2 \beta^3 N-446 A^2 \beta^2 N-366 \beta^6 N^3-306 A^2 \beta^3-52 \beta^6 N^2-198 \beta^2 N^2 A^2+32 \beta^4 N^4 A^2+232 \beta^4 N^3 A^2-192 \beta^3 N^3 A^2-134 \beta^4 N^2 A^2+9 A^6-36 \beta^6-54 A^4 \beta^2+78 A^4 \beta+81 A^2 \beta^4+104 \beta^2 N^3 A^2+80 \beta^7 N^3+12 \beta^7 N^2-36 N^4 \beta^3 A^2-48 \beta^5 N^4 A^2-116 \beta^5 N^3 A^2+21 \beta^5 N^2 A^2+12 N^5 \beta^5 A^2-42 \beta^3 N^2 A^4+4 \beta^5 N^6 A^2-16 \beta^4 N^5 A^2+12 \beta^3 N^4 A^4+24 \beta^3 N^3 A^4+228 \beta^6 N-318 \beta^4 N A^2+90 A^4 \beta^2 N+66 \beta^2 N^2 A^4-24 \beta^2 N^3 A^4-66 \beta N A^4+81 \beta^5 A^2 N-36 \beta^7 N-54 \beta^3 A^4 N+9 N A^6 \beta-42 \beta N^2 A^4+9 \beta N^2 A^6) \Big[ 2 (A-\beta+1) (A+\beta-1) (A-1+2 \beta) (A-2+\beta) (A+2-\beta) (A+1-2 \beta) \Big]^{-1} $
\smallskip\\
\end{center}
Let us now check this prediction for particular values of $N$ and $\beta$.
\pagebreak

\subsubsection*{The case of $(N,\beta) = (1,1)$}

For natural $\mu$ one has:
\be
J^{(exp)}_2\big( A,\mu,1,1 \big) = \dfrac{\mu (-1+2 \mu) (\mu^4+2 \mu^3+5 \mu^2+6 \mu^2 A^2+16 \mu-18 \mu A^2+12-33 A^2+9 A^4)}{(A+1+\mu) (A-1-\mu) (A-2-\mu) (A+2+\mu)}
\ee
\smallskip\\
Analytically continuing to $\mu = -\beta N = -1$, one finds
\be
j^{(exp)}_2\big( A,1,1 \big) = 27 = j^{(theor)}_2\big( A,1,1 \big)
\ee
\smallskip\\
This provides a check of eq. (\ref{MainIdent}) at level 2 and in the case
of $(N,\beta) = (1,1)$.

\subsubsection*{The case of $(N,\beta) = (2,1)$}

For natural $\mu$ one has:
\begin{center}
$J^{(exp)}_2\big( A,\mu,2,1 \big) = \Big( 81 A^2-54 A^4+9 A^6-36+174 \mu A^2-408 \mu A^4+54 \mu A^6+84 \mu+968 \mu^2 A^2-867 \mu^2 A^4+72 \mu^2 A^6+207 \mu^2+1152 \mu^3 A^2-402 \mu^3 A^4-102 \mu^3+395 \mu^4 A^2-24 \mu^4 A^4-242 \mu^4-30 \mu^5 A^2+24 \mu^5-40 \mu^6 A^2+79 \mu^6-6 \mu^7-8 \mu^8 \Big) \Big[ (A+2+\mu) (A-2-\mu) (A-3-\mu) (A-1-\mu) (A+1+\mu) (A+3+\mu) \Big]^{-1}$
\smallskip\\
\end{center}
Analytically continuing to $\mu = -\beta N = -2$, one finds
\be
j^{(exp)}_2\big( A,2,1 \big) = \dfrac{9(21 A^4 + 35 A^2 - 64)}{A^2(A^2-1)} = j^{(theor)}_2\big( A,2,1 \big)
\ee
\smallskip\\
This provides a check of eq. (\ref{MainIdent}) at level 2 and in the case
of $(N,\beta) = (2,1)$.

\subsubsection*{The case of $(N,\beta) = (3,1)$}

For natural $\mu$ one has:
\begin{center}
$J^{(exp)}_2\big( A,\mu,3,1 \big) = \Big( 2700 A^2-1755 A^4+135 A^6-1728 \mu+12888 \mu A^2-5391 \mu A^4+297 \mu A^6-13356 \mu^2+21942 \mu^2 A^2-5157 \mu^2 A^4+162 \mu^2 A^6-16632 \mu^3+12978 \mu^3 A^2-1503 \mu^3 A^4-5787 \mu^4+2061 \mu^4 A^2-54 \mu^4 A^4+861 \mu^5-333 \mu^5 A^2+657 \mu^6-90 \mu^6 A^2+3 \mu^7-18 \mu^8 \Big) \Big[ (A+3+\mu) (A-3-\mu) (A-4-\mu) (A-2-\mu) (A+2+\mu) (A+4+\mu) \Big]^{-1}$
\smallskip\\
\end{center}
Analytically continuing to $\mu = -\beta N = -3$, one finds
\be
j^{(exp)}_2\big( A,3,1 \big) = \dfrac{54(13 A^6 + 78 A^4 - 123 A^2+192)}{A^2(A^2-1)^2} = j^{(theor)}_2\big( A,3,1 \big)
\ee
\smallskip\\
This provides a check of eq. (\ref{MainIdent}) at level 2 and in the case
of $(N,\beta) = (3,1)$.

\subsubsection*{The case of $(N,\beta) = (1,2)$}

For natural $\mu$ one has:
\be
J^{(exp)}_2\big( A,\mu,1,2 \big) = \dfrac{\mu(-1+2 \mu)(\mu^4+2 \mu^3+5 \mu^2+6 \mu^2 A^2+16 \mu-18 \mu A^2+12-33 A^2+9 A^4)}{(A+1+\mu) (A-1-\mu) (A-2-\mu) (A+2+\mu)}
\ee
\smallskip\\
Analytically continuing to $\mu = -\beta N = -2$, one finds
\be
j^{(exp)}_2\big( A,1,2 \big) = \dfrac{90(A^2+3)}{A^2-1} = j^{(theor)}_2\big( A,1,2 \big)
\ee
\smallskip\\
This provides a check of eq. (\ref{MainIdent}) at level 2 and in the case
of $(N,\beta) = (1,2)$. In fact, the $\mu$-dependent answer coincides completely
with the $(N,\beta) = (1,1)$ case, i.e. it does not depend on $\beta$.
This property is specific for $N = 1$, where the Van-der-Monde factor actually does
not contribute to the partition function.

\subsubsection*{The case of $(N,\beta) = (2,2)$}

For natural $\mu$ one has:
\begin{center}
$J^{(exp)}_2\big( A,\mu,2,2 \big) = \Big( -576+846 A^2-324 A^4+54 A^6-192 \mu+198 \mu A^2-1332 \mu A^4+126 \mu A^6+1298 \mu^2+2000 \mu^2 A^2-1578 \mu^2 A^4+72 \mu^2 A^6+402 \mu^3+2328 \mu^3 A^2-522 \mu^3 A^4-876 \mu^4+722 \mu^4 A^2-24 \mu^4 A^4-228 \mu^5-6 \mu^5 A^2+162 \mu^6-40 \mu^6 A^2+18 \mu^7-8 \mu^8 \Big) \Big[ (A+3+\mu) (A-3-\mu) (A-4-\mu) (A-1-\mu) (A+1+\mu) (A+4+\mu) \Big]^{-1}$
\smallskip\\
\end{center}
Analytically continuing to $\mu = -\beta N = -4$, one finds
\be
j^{(exp)}_2\big( A,2,2 \big) = \dfrac{54(-2800-1663 A^2+130 A^4+13 A^6)}{A^2(A^2-1)(A^2-9)} = j^{(theor)}_2\big( A,2,2 \big)
\ee
\smallskip\\
This provides a check of eq. (\ref{MainIdent}) at level 2 and in the case
of $(N,\beta) = (2,2)$.

\subsubsection*{The case of $(N,\beta) = (3,2)$}

For natural $\mu$ one has:
\begin{center}
$J^{(exp)}_2\big( A,\mu,3,2 \big) = \Big( -3240+36666 A^2-13860 A^4+594 A^6-48276 \mu+121329 \mu A^2-23994 \mu A^4+621 \mu A^6-201978 \mu^2+133794 \mu^2 A^2-13122 \mu^2 A^4+162 \mu^2 A^6-177237 \mu^3+52956 \mu^3 A^2-2367 \mu^3 A^4-50814 \mu^4+6174 \mu^4 A^2-54 \mu^4 A^4-66 \mu^5-441 \mu^5 A^2+1746 \mu^6-90 \mu^6 A^2+75 \mu^7-18 \mu^8 \Big) \Big[ (A+5+\mu) (A-5-\mu) (A-6-\mu) (A-3-\mu) (A+3+\mu) (A+6+\mu) \Big]^{-1}$
\smallskip\\
\end{center}
Analytically continuing to $\mu = -\beta N = -6$, one finds
\be
j^{(exp)}_2\big( A,3,2 \big) = \dfrac{180(15 A^6+550 A^4-453 A^2-21168)}{A^2(A^2-1)^2(A^2-9)^2} = j^{(theor)}_2\big( A,3,2 \big)
\ee
\smallskip\\
This provides a check of eq. (\ref{MainIdent}) at level 2 and in the case
of $(N,\beta) = (3,2)$.

\subsubsection*{The case of $(N,\beta) = (1,3)$}

For natural $\mu$ one has:
\be
J^{(exp)}_2\big( A,\mu,1,3 \big) = \dfrac{\mu(-1+2 \mu)(\mu^4+2 \mu^3+5 \mu^2+6 \mu^2 A^2+16 \mu-18 \mu A^2+12-33 A^2+9 A^4)}{(A+1+\mu) (A-1-\mu) (A-2-\mu) (A+2+\mu)}
\ee
\smallskip\\
Analytically continuing to $\mu = -\beta N = -3$, one finds
\be
j^{(exp)}_2\big( A,1,3 \big) = \dfrac{63(3 A^4 + 25 A^2 + 12)}{(A^2-1)(A^2-4)} = j^{(theor)}_2\big( A,1,3 \big)
\ee
\smallskip\\
This provides a check of eq. (\ref{MainIdent}) at level 2 and in the case
of $(N,\beta) = (1,3)$. Again, the $\mu$-dependent answer coincides completely with the $(N,\beta) = (1,1)$ case, because of $\beta$-independence of the partition function.

\subsubsection*{The case of $(N,\beta) = (2,3)$}

For natural $\mu$ one has:
\begin{center}
$J^{(exp)}_2\big( A,\mu,2,3 \big) = \Big( -3600+4095 A^2-630 A^4+135 A^6-3240 \mu-2334 \mu A^2-2544 \mu A^4+198 \mu A^6+4705 \mu^2+1880 \mu^2 A^2-2397 \mu^2 A^4+72 \mu^2 A^6+3702 \mu^3+3792 \mu^3 A^2-642 \mu^3 A^4-1666 \mu^4+1157 \mu^4 A^2-24 \mu^4 A^4-864 \mu^5+18 \mu^5 A^2+209 \mu^6-40 \mu^6 A^2+42 \mu^7-8 \mu^8 \Big) \Big[ (A+4+\mu) (A-4-\mu) (A-5-\mu) (A-1-\mu) (A+1+\mu) (A+5+\mu) \Big]^{-1}$
\smallskip\\
\end{center}
Analytically continuing to $\mu = -\beta N = -6$, one finds
\be
j^{(exp)}_2\big( A,2,3 \big) = \dfrac{27(57 A^6+1330 A^4-45927 A^2-425860)}{(A^2-1)(A^2-4)(A^2-25)} = j^{(theor)}_2\big( A,2,3 \big)
\ee
\smallskip\\
This provides a check of eq. (\ref{MainIdent}) at level 2 and in the case
of $(N,\beta) = (2,3)$. Together with the previous checks, this provides a rather
firm evidence of validity of (\ref{MainIdent}) at levels 1, 2 and at finite $N$,
what is the main claim of the present paper.

\section{Discussion
\label{conc}}

In this paper we explicitly demonstrated that if the Dotsenko-Fateev
(DF) $\beta$-ensemble of \cite{DV,MMS1,MMS2} is defined as in
refs.\cite{DV,MY} and the screening integration contour is
chosen along the A-cycle, then it indeed
reproduces the 1-point toric conformal
block. We made an explicit check in the first two orders of
$q$-expansion, but in this field this is well-known to be sufficient
to get rid of any possible doubts. The two crucial features of this
{\it new} DF representation for {\it generic} conformal blocks,
which makes it different from the original DF formulas for minimal
models and alike \cite{DF,Fel}, are

(i) the use of a single (rather than two)
screening charges, and

(ii) the use
of contour integrals along {\it open}
(not obligatory/necessarilly {\it closed})
contours.
\\
The procedure also includes an analytical continuation
in $\alpha$-parameters, but since every term
of the $q$-expansion is a {\it rational} function
of $\alpha$'s, this is not really a problem,
as long as the conformal block is regarded as no more
than a formal series (\ref{cobloexp}) in $q$.
The main difficulty at this level of consideration
is specification of the integration contours.

To understand the difference between the present paper
and \cite{DV,MY}, one should remember that there
are three levels of accuracy in the definition of
the DF $\beta$-ensembles \cite{MSdv}.

At the first level, one simply reproduces the Seiberg-Witten (SW)
differential and the SW prepotential \cite{SWth,GKMMM,Cal}. For this purpose
one just needs to take the $g_s = \sqrt{-\epsilon_1\epsilon_2}
\rightarrow 0$ limit of the $\beta$-ensemble in the Dijkgraaf-Vafa
(DV) phase \cite{DVth}, when the eigenvalues are concentrated within
the cuts around extrema of the matrix-model potential $W$
\cite{multicut}. This limit is controlled by the quasiclassical
approximation, it does not depend on the choice of integration
contours (as soon as all extrema are within the integration
domains), and it directly describes the SW prepotential as the
$\beta$-ensemble free energy in terms of the spectral SW curve. The
curve is seen already in the expression for the first resolvent
$\rho^{(0|1)}(z) = \ \left<\tr \frac{1}{\phi-z}\right>$. Note that
$\beta = b^2 = -\epsilon_1/\epsilon_2$ can already be arbitrary,
only $g_s$ is kept small. Most of the papers on matrix models and
$\beta$-ensembles in the context of the AGT relation, including
\cite{MY}, are devoted to {\it this} level of consideration.

At the next level, one restores all terms of the
genus-expansion for the matrix model ($\beta$-ensemble)
free energy. This can be most effectively done by
the old resolvent techniques, which was recently
revived under the name of the AMM-EO topological recursion
\cite{TR}: as a general method of constructing
the hierarchy of resolvents,
implied by the Virasoro or $W$-constraints,
starting from an arbitrary spectral curve defined as a covering.
In application to the DV phases this method provides
the free energy as a series in powers of $g_s$ and rescaled
multiplicities $S_\nu = g_sN_\nu$.
In the original DV theory these
series in $S_q$ are also known as the CIV-DV prepotentials
(with higher genus corrections).
In application to the DF integrals this approach provides
expressions which are exact in $q$-parameters.

However, even this level of consideration is insufficient
for the study of the AGT relations.
The problem is that the conformal blocks are usually known in
the form of $q$-series (\ref{cobloexp}),
but instead each coefficient ${\cal B}_k$ is a {\it rational}
function of the dimensions, and, thus, of the multiplicities $N_\nu$.
This means that the CIV-DV series in $S_\nu$ should be exactly
summed up before they can be compared with the known
expressions for generic conformal blocks.
In other words \cite{MSdv}, from the point of view of
the current AGT studies the topological recursion
provides an excessive information about the $q$-dependencies,
but insufficient information about the $N_\nu$-dependencies.
Therefore, at the third level of accuracy, one needs to
specify {\it exactly} the integration contours in the DF integrals
\cite{MMS1,MMS2},
then these integrals can be {\it evaluated explicitly}
-- at least, after the $q$-expansion is performed and
integrals belong to the (slightly extended)
Euler-Selberg family \cite{MMMor}.
Expanding the answers in powers of $N_\nu$, one returns back
to the CIV-DV potentials.
This comparison of exact integrals and the CIV-DV expansions
was explicitly performed in \cite{MSdv} for the standard
simplest example of the 4-point conformal block, but there
the guess of \cite{MMS2} was used for the proper choice
of the contours. Unfortunately, there is still no clear
idea of how the contours should be selected for the conformal
blocks on higher genus Riemann surfaces.
In the present paper, we checked the new {\it guess}
of \cite{DV,MY}: that for $g=1$ an additional contour
should be the $A$-cycle, provided the matrix model potential
is {\bf modified {\it al la} \cite{DV},
by adding the shift along the Jacobian}.

A motivation for this shift comes from calculating
the free parameters in the AGT relation.
A genus $g$ conformal block with $n$ external
lines (punctures) has $3g-3+2n$
free parameters: the dimensions of fields on
$n$ external and $3g-3+n$ internal lines
of the corresponding Feynman diagram.
This same number of parameters should be of course
present in the DF representation of the conformal block.
At the {\it first} level in above classification,
this means that the number $P$ of parameters in
the matrix model potential $W(z)$ and the number $Z$ of
zeroes of $dW(z)$ should sum up to
\be
P + Z -1 = 3g-3+2n
\ee
Unity is subtracted from the l.h.s. because of
the "conservation law"
\be
\left(
\sum_{i=1}^n \alpha_i + b\sum_{\nu=1}^{2g+n-2} N_\nu
\right) =
\left({1\over b} - b\right)(g-1)
\label{conslaw}
\ee
which is a characteristic feature of the DF integrals.
The number of zeroes of the differential with
$n$ poles on the genus $g$ Riemann surface
(not obligatory single-valued)
is given by the Riemann-Roch theorem:
it is $Z=2g+n-2$.
Thus, $P$ should be equal to $P=n+g$.
Of these $n$ are the $\alpha$-parameters of the $n$ external
lines, and $g$ should be looked for somewhere else.
A suggestion of \cite{DV} was simply to add an arbitrary
linear combination of holomorphic differentials to $dW(z)$:
\be
dW = 2b\sum_{i=1}^n \alpha_i d\log E(z,x_i)
+ \sum_{k=1}^g p_k\omega_k
\label{dW}
\ee
where
$E(z,x) = \frac{\theta_*(\vec z-\vec x)}{\nu_*(z)\nu_*(x)}$
is the prime-form,
$x$'s are positions of the $n$ punctures
and $\omega_k$ are the $g$ holomorphic differentials.
The $n+g$ free parameters are $\alpha_i$ and $p_k$.
In \cite{MY} it was checked that, for $g=1$, this prescription
reproduces the relevant SW curve,
and in the present paper we checked much more: that
the relevant conformal block is also reproduced.

For this, however, we had to specify the DF integral in
more details. In the DF representation, the multiplicities $N_\nu$
are the multiplicities of different integrals of the screening
operator $V_b = e^{ib\phi}$, and to fully define the integral
one should explicitly specify the $\#(\nu) = Z = 2g+n-2$
integration contours.
According to \cite{MMS2}, for genus $g=0$
these $n-2$  contours connect the
pairs of external lines: the first and the second,
the first and the third and so on, up to the $(n-1)$-st one,
while the $n$-th puncture is always at infinity:
$dW^{(g=0)} = \sum_{i=1}^{n-1}\frac{2b\alpha_i}{z - x_i}\ dz$
and $\alpha_n$ is defined from the conservation law (\ref{conslaw}).
For $g>0$ the choice of contours is far more obscure.
For $g=1$ the suggestion of \cite{DV,MY} seems to be:
$n-1$ contour between the pairs of the punctures and one
additional contour along the $A$-cycle.
In this paper we checked that this indeed works for $n=1$
and there are few doubts that this will be true for all $n>1$,
though this remains to be checked as well.
Unfortunately, this sheds no light on possible choices
of contours for $g\geq 2$: there are still $n-1$ "obvious"
contours between the punctures, but it is not very clear
what the remaining $2g-1$ are going to be.
This puzzle remains to be resolved.

It deserves noting that from the point of view of DF
representation there could be another possibility,
mentioned in \cite{MMS2}.
Instead of adding $p_k$ parameters to (\ref{dW}),
one could simply choose $3g-1$ additional contours,
so that multiplicities of the corresponding
integrations  complemented the $n$ external
$\alpha$-parameters and $n-1$ multiplicities of
"obvious contours" between the punctures to
match the number of parameters in conformal block:
\be
3g-1 + n + (n-1) -1 = 3g+2n-3
\ee
As usual, unity is subtracted from the l.h.s.
because of the conservation law (\ref{conslaw}),
where sum over $\nu$ this times goes from $1$ to
$3g+n-2$.
The point of \cite{MMS2} was that $3g-1$ is a nice
number: it is the quantity of {\it non-homotopic}
closed contours on the genus $g$ Riemann surface
(which for $g>1$ exceeds the number $2g$ of
non-homological $A$ and $B$ cycles).
Unfortunately, we did not manage to make this
prescription working even for $(g,n)=(1,1)$,
what, perhaps, is not a surprise because it was not
made consistent with the first (quasiclassical) level
of the DV description.
Still, while the problem of contour choice remains
open, one should not full neglect this alternative
possibility. (For example, one can use all the non-homotopic
contours besides $B$-cycles, getting $2g-1=3g-1-g$
closed contours...)

Of course, for any consideration of the AGT relations
at higher genera, $g>1$, one should also deduce, at least,
the first terms of $q$-expansions of higher genus
conformal blocks from representation theory.
This is a straightforward, but tedious calculation
of its own value, which is not yet reported in the
literature.

\section*{Acknowledgements}

Our work is partly supported by Russian Federal Nuclear Energy Agency,
Russian Federal Agency for Science and Innovations under contract
14.740.11.0081, by RFBR grants 10-02-00509- 
(A.Mir.), and 10-02-00499 (A.Mor. \& Sh.Sh.), by joint grants 09-02-90493-Ukr,
09-01-92440-CE, 09-02-91005-ANF, 10-02-92109-Yaf-a.
The work of A.Morozov is also supported in part by CNRS and that of
Sh.Shakirov by Moebius Contest Foundation for
Young Scientists.

\end{document}